\documentclass[letterpaper]{jpconf}

\usepackage{graphicx}

\begin{document}

\title{The equation of state of neutron star matter and the symmetry energy}
\author{Stefano Gandolfi}
\address{Theoretical Division, Los Alamos National Laboratory
Los Alamos, NM 87545, USA}
\ead{stefano@lanl.gov}

\begin{abstract}
We present an overview of microscopical calculations of the Equation
of State (EOS) of neutron matter performed using Quantum Monte Carlo
techniques.  We focus to the role of the model of the three-neutron
force in the high-density part of the EOS up to a few times the saturation
density.  We also discuss the interplay between the symmetry energy and
the neutron star mass-radius relation.

The combination of theoretical models of the EOS with recent neutron
stars observations permits us to constrain the value of the symmetry energy
and its slope. We show that astrophysical observations are starting
to provide important insights into the properties of neutron star matter.

\end{abstract}

\section{Introduction}

The knowledge of the Equation of State (EOS) of pure neutron matter
is an important bridge between the symmetry energy, and
neutron star properties.  The symmetry energy $E_s$ is the difference of
nuclear matter and neutron matter energy, it gives the energy cost of the
isospin-asymmetry in the homogeneous nucleonic matter.  In the last few
years the study of $E_s$ has received considerable attention (see for example
Ref.~\cite{Tsang:2012} for a recent experimental/theoretical review).

The role of the symmetry energy is
essential to understand the mechanism of stability of very-neutron
rich nuclei, but it is also related to many phenomena occurring in
neutron stars.
The stability of matter inside neutron stars is very sensitive to
$E_s$ and its first derivative.  
Around saturation density, neutrons tend to decay to protons through the
$\beta$-decay, and the cooling of neutron stars is strongly connected to
the proton/neutron ratio as a function of the density.  This ratio is
mainly governed by the behavior of $E_s$ as a function of the density.

The inner crust of neutron stars, where the density is a fraction
of nuclear densities, is mostly composed of neutrons surrounding
a matter made of extremely-neutron rich nuclei that, depending
on the density, may exhibit very different phases and properties.
The extremely rich phase diagram of the neutron crustal matter is
strongly related to the role of $E_s$.  For example it governs the
phase-transition between the crust and the core~\cite{Newton:2011}
and $r$-mode instability~\cite{Wen:2012,Vidana:2012}.  

The study of the EOS is particularly difficult because neutron
matter is one of the most strongly-interacting fermionic systems.
Neutron matter is often modeled by density functionals. Traditional
Skyrme models (see for example Ref. \cite{Stone:2007} and references
therein) and relativistic mean-field models (see for example
Refs.~\cite{Fattoyev:2010,Fattoyev:2012}) are two general classes of
density functional theories.

Recently, a new methods based on microscopic nuclear Hamiltonians
obtained from chiral effective field theories have been proposed. These
nuclear forces are obtained following a systematic expansion in terms of
momenta of the relevant degrees of freedom, and fit the nucleon-nucleon
scattering data~\cite{Entem:2003}. The nuclear Hamiltonian is then
adjusted using renormalization group techniques to make the calculation
perturbative~\cite{Hebeler:2010}. New induced terms generated trough the
renormalization scheme have not yet been included though. These effects,
as well as non-perturbative effects of three- and four-body forces,
could be important~\cite{Roth:2011}.

The third class of these calculations uses nuclear potentials,
like Argonne and Urbana/Illinois forces, that reproduces
two-body scattering and properties of light nuclei with very high
precision~\cite{Wiringa:1995,Pieper:2001}. In the latter case,
the interaction has small non-local terms, giving the potentials
a hard core. In this case calculations can be performed in a
non-perturbative framework, and the strong correlations are
solved by using correlated wave functions.  In this paper we
present results based on quantum Monte Carlo (QMC) methods. QMC
methods have proven to be a very powerful tool to accurately
study properties of light nuclei~\cite{Pudliner:1997,Pieper:2008}
and nuclear matter~\cite{Gandolfi:2010} in a very similar way.
They provide the unique technique to date to consistently study
nuclear systems of different kind, both inhomogeneous and homogeneous
matter, with the same accuracy and using the same Hamiltonians.
Other techniques have been used to study nuclear and neutron matter
and the symmetry energy based on Brueckner-Hartree-Fock theory (see for
example Ref.~\cite{Vidana:2009} and references therein).

\section{Nuclear Hamiltonian and Quantum Monte Carlo method}

In our model, neutrons are non-relativistic point-like particles
interacting via two- and three-body forces:
\begin{eqnarray}
H = \sum_{i=1}^A\frac{p_i^2}{2m} + \sum_{i<j}v_{ij}+\sum_{i<j<k}v_{ijk} \,.
\end{eqnarray}
The two body-potential that we use is the Argonne
AV8'~\cite{Wiringa:2002}, that is a simplified form of the Argonne
AV18~\cite{Wiringa:1995}. Although simpler to use in QMC calculations,
the AV8' provides almost the same accuracy as AV18 in fitting NN
scattering data.  The three-body force is not as well constrained
as the NN interaction, but its inclusion in realistic nuclear
Hamiltonians is important to correctly describe the binding energy
of light nuclei~\cite{Pieper:2001}. Note that this is not necessary
if phenomenological forces, like Skyrme or Gogny, or methods based on
Relativistic Mean-Field theory are used.

The Urbana IX (UIX) three-body force has been originally proposed in
combination with the Argonne AV18 and AV8'~\cite{Pudliner:1995}. Although
it slightly underbinds the energy of light nuclei, it has been
extensively used to study the equation of state of nuclear and neutron
matter~\cite{Akmal:1998,Gandolfi:2009,Gandolfi:2012}.  The Illinois
forces have been introduced to improve the description of both
ground- and excited-states of light nuclei, showing an excellent
accuracy~\cite{Pieper:2001,Pieper:2008}, but it produces an unphysical
overbinding in pure neutron systems~\cite{Sarsa:2003}.  In this paper we
shall present a study of the neutron matter EOS based on different models
of three-neutron force giving specific values of the symmetry energy.

We solve the many-body ground-state using the Auxiliary Field
Diffusion Monte Carlo (AFDMC) originally introduced by Schmidt and
Fantoni~\cite{Schmidt:1999}. The main idea of QMC methods is to evolve
a many-body wave function in imaginary-time:
\begin{equation}
\Psi(\tau)=\exp[-H\tau]\Psi_v \,,
\end{equation}
where $\Psi_v$ is a variational ansatz, and $H$ is the Hamiltonian
of the system.  In the limit of $\tau\rightarrow\infty$, $\Psi$
approaches the ground-state of $H$.  The evolution in imaginary-time
is performed by sampling configurations of the system using Monte
Carlo techniques, and expectation values are evaluated over
the sampled configurations.  For more details see for example
Refs.~\cite{Pudliner:1997,Gandolfi:2009}.

The Green's Function Monte Carlo (GFMC) provided to be extremely accurate
to study properties of light nuclei. The variational wave function
includes all the possible spin/isospin states of nucleons and it
provides a good variational ansatz to start the projection in the
imaginary-time. The exponential growing of this states limits the
calculation to the $^{12}C$~\cite{Pieper:2005}.
The AFDMC method does not explicitly include all the spin/isospin states in the wave
function, but they are instead sampled using the Hubbard-Stratonovich
transformation. The calculation can be then extended up to many neutrons,
making the simulation of homogeneous matter possible.  The AFDMC has proven to
be very accurate when compared to GFMC calculation of energies of neutrons
confined in an external potential~\cite{Gandolfi:2011}.

\section{Symmetry energy}

The symmetry energy is defined as the difference between
pure neutron matter and symmetric nuclear matter. The energy of nuclear
matter is often expressed as an expansion in even powers of the 
isospin-asymmetry
\begin{equation}
E(\rho,x)=E_0(\rho)+E_s^{(2)}(\rho)(1-2x)^2+E_s^{(4)}(1-2x)^4+\dots \,,
\end{equation}
where $E$ is the energy per particle, $x=\rho_p/(\rho_p+\rho_n)$ is
the proton fraction, $\rho$ is the density of the system, $E_s^{(2n)}$
are coefficients multiplying the isospin asymmetry terms $(1-2x)^{2n}$, and
$E_0(\rho)=E(\rho,x=0.5)$ is the energy of symmetric nuclear matter.  
The symmetry
energy $E_s$ is given by
\begin{equation}
E_s(\rho)=E(\rho,0)-E_0(\rho) \,.
\end{equation}
The energy at saturation of symmetric nuclear matter extrapolated
from the binding energy of heavy nuclei is $E(\rho_0)=-16$ MeV, where
$\rho_0=0.16$ fm$^{-3}$ is the saturation density.  The symmetry energy
around saturation $\rho_0$ can be expanded as
\begin{equation}
E_s(\rho)\Big|_{\rho_0}=E_{sym}+\frac{L}{3}\frac{\rho-\rho_0}{\rho_0}+\dots \,,
\label{eq:lvsesym}
\end{equation}
where $L$ is related to the slope of $E_s$.
By combining the above equations, we can easily relate the symmetry
energy to the EOS of pure neutron matter at density close to $\rho_0$.

\section{Results}

In this section we present QMC results for pure neutron matter.
There are several reasons to focus on pure neutron matter. First, the
three-body interaction is non-zero only in the $T=3/2$ isospin-channel ($T$ is the
total isospin of three-nucleons), while in the presence of protons there
are also contributions in $T=1/2$. The latter term is the dominant one in
nuclei, and only weakly accessible by studying properties of nuclei.
Second, the EOS of pure neutron matter is closely related to the structure of
neutron stars.

\begin{figure}
\begin{center}
\includegraphics[width=0.7\textwidth]{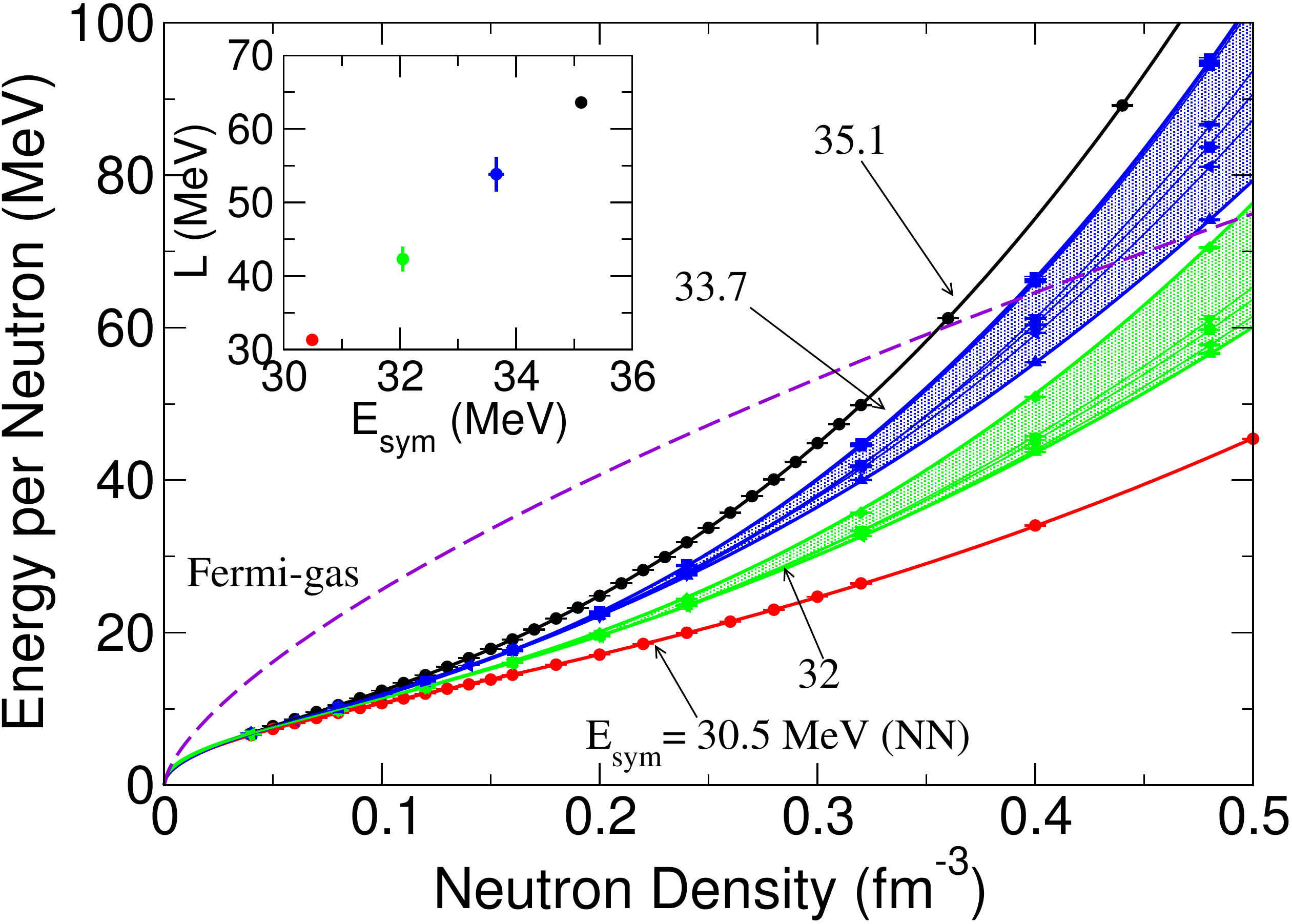}
\end{center}
\caption{The QMC equation of state of neutron matter for various Hamiltonians.
The red (lower) curve is obtained by including the NN (Argonne AV8') alone in the
calculation, and the black one is obtained by adding the Urbana IX three-body force.
The green and blue bands correspond to EOS giving the same $E_{sym}$ (32 and 33.7 MeV
respectively), and are obtained by using several models of three-neutron force.
In the inset we show the value of $L$ as a function of $E_{sym}$ obtained by fitting
the EOS.
The figure is taken from Ref.~\cite{Gandolfi:2012}.
}
\label{fig:eos}
\end{figure}

We present several EOS obtained using different models of three-neutron
force in Fig.~\ref{fig:eos}. The two solid lines correspond to the EOS
calculated using the NN potential alone and including the UIX three-body
force~\cite{Pudliner:1995}. The effect of using different models of three-neutron force
is clear in the two bands, where the high density behavior is showed up
to about $3\rho_0$.  At such high density, the various models giving the
same symmetry energy at saturation produce an uncertainty to the EOS of
about 20 MeV.
The EOS obtained using QMC can be conveniently fit using the following
functional~\cite{Gandolfi:2009}:
\begin{equation}
E(\rho)=a\,\left(\frac{\rho}{\rho_0}\right)^\alpha
+b\,\left(\frac{\rho}{\rho_0}\right)^\beta \,,
\label{eq:enefunc}
\end{equation}
where $E$ is the energy per neutron, $\rho_0=0.16$ fm$^{-3}$, 
and $a$, $b$, $\alpha$ and $\beta$ are free parameters.
The parametrizations of the EOS obtained from different nuclear Hamiltonians
is given in Ref.~\cite{Gandolfi:2012}.

At $\rho_0$ symmetric nuclear matter saturates, and we can extract the value of $E_{sym}$ and $L$
directly from the pure neutron matter EOS. The result of fitting
Eq.~\ref{eq:lvsesym} to the pure neutron matter EOS is shown in the
inset of Fig.~\ref{fig:eos}.  The error bars are obtained by taking the
maximum and minimum value of $L$ for a given $E_{sym}$, and the curves
obtained with NN and NN+UIX are thus without error bars. From the plot it
is clear that within the models we consider, the correlation between $L$
and $E_{sym}$ is linear and quite strong.

When the EOS of the neutron matter has been specified, the structure 
of an idealized spherically-symmetric neutron star model can be calculated 
by integrating the Tolman-Oppenheimer-Volkoff (TOV) equations:
\begin{equation}
\frac{dP}{dr}=-\frac{G[m(r)+4\pi r^3P/c^2][\epsilon+P/c^2]}{r[r-2Gm(r)/c^2]} \,,
\label{eq:tov1}
\end{equation}
\begin{equation}
\frac{dm(r)}{dr}=4\pi\epsilon r^2 \,,
\label{eq:tov2}
\end{equation}
where $P=\rho^2(\partial E/\partial\rho)$ and $\epsilon=\rho(E+m_N)$
are the pressure and the energy density, $m_N$ is the neutron mass,
$m(r)$ is the gravitational mass enclosed within a radius $r$, and $G$
is the gravitational constant. The solution of the TOV equations for a
given central density gives the profiles of $\rho$, $\epsilon$ and $P$ as
functions of radius $r$, and also the total radius $R$ and mass $M=m(R)$.
The total radius $R$ is given by the condition $P(R)=0$.

\begin{figure}
\begin{center}
\includegraphics[width=0.7\textwidth]{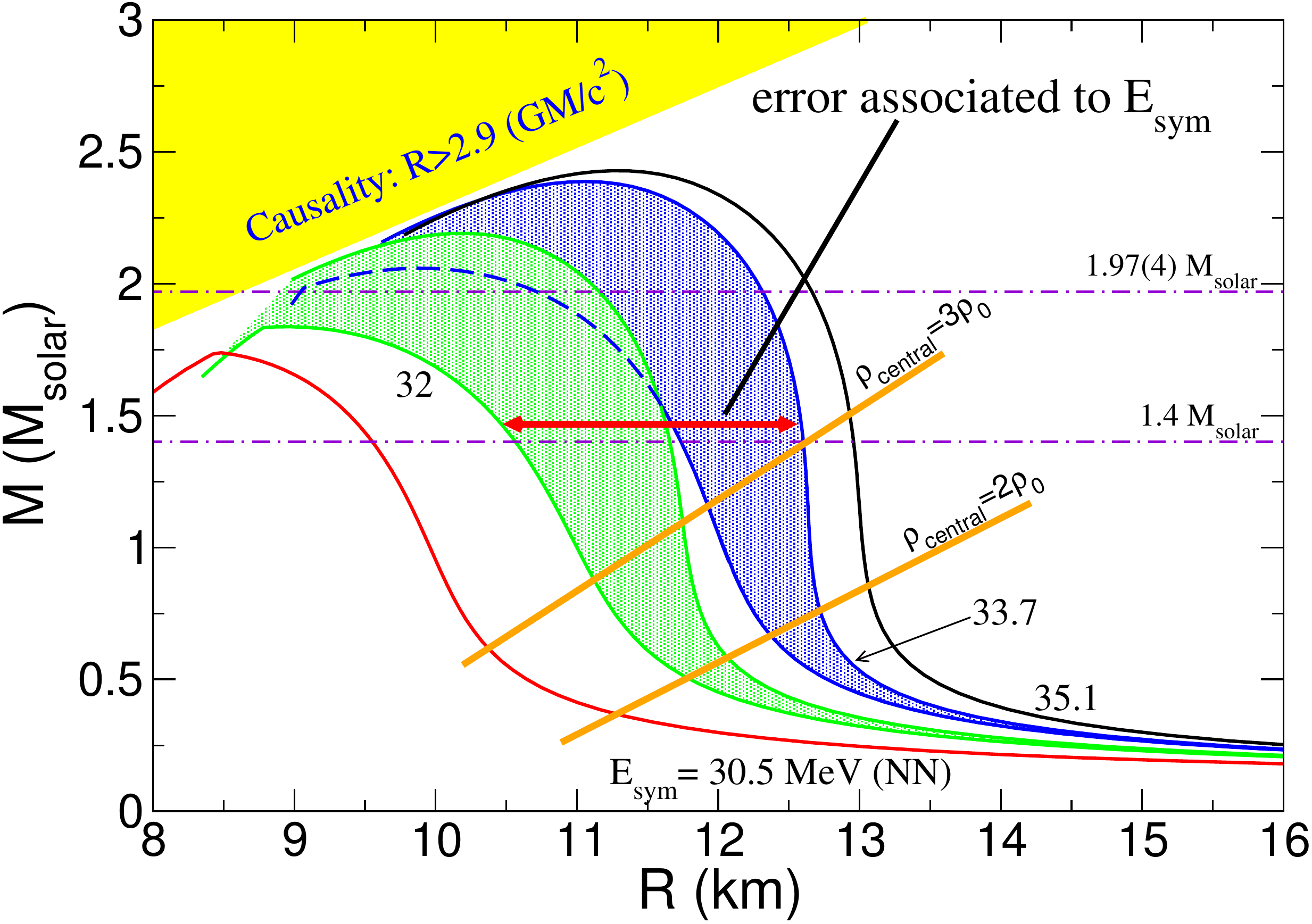}
\end{center}
\caption{The mass-radius relation of neutron stars obtained from
the EOS calculated using QMC. The various colors represent the $M-R$
result obtained from the corresponding EOS obtained as described in
Fig.~\ref{fig:eos}.  The two horizontal lines show the value of $M=1.4$
and 1.97(4)$M_{solar}$~\cite{Demorest:2010}.
The figure is taken from Ref.~\cite{Gandolfi:2012}.
}
\label{fig:nstar}
\end{figure}

The mass of a neutron star as a function of its radius is shown in
Fig.~\ref{fig:nstar}. The two bands correspond to the result obtained
using the two sets of EOS giving the same value of $E_{sym}$ indicated
in the figure. As in the case of the EOS, it is clear that the main 
source of uncertainty to the radius of a neutron star with $M=1.4M_{solar}$ 
is due to the uncertainty of $E_{sym}$ rather than the model of 
the three-neutron force.
It has to be noted that we have used the EOS of pure neutron matter
without imposing the $\beta$-equilibrium, so in our model we don't have
protons. However the addition of a small proton fraction would change
the radius $R$ only slightly~\cite{Gandolfi:2010,Akmal:1998} smaller
than other uncertainties in the EOS that we have discussed.

The EOS of neutron matter and its properties can also be
extracted from astrophysical observations~\cite{Steiner:2010}. By
combining the Bayesian analysis with the model of neutron matter of
Eq.~\ref{eq:enefunc} it is possible to compare the QMC prediction with
observations~\cite{Steiner:2012}, and to extract $E_{sym}$ and $L$:
\begin{equation}
E_{sym}=a+b+16 \,,\quad L=3\,(a\alpha+b\beta) \,,
\end{equation}
and from neutron stars we obtain the constraints
$31.2<E_{sym}<34.3$ MeV and $36.6<L<55.1$ MeV~\cite{Steiner:2012} (at the 
$2$-$\sigma$ confidence level) in agreement
with the QMC predictions.

\begin{figure}
\begin{center}
\includegraphics[width=0.7\textwidth]{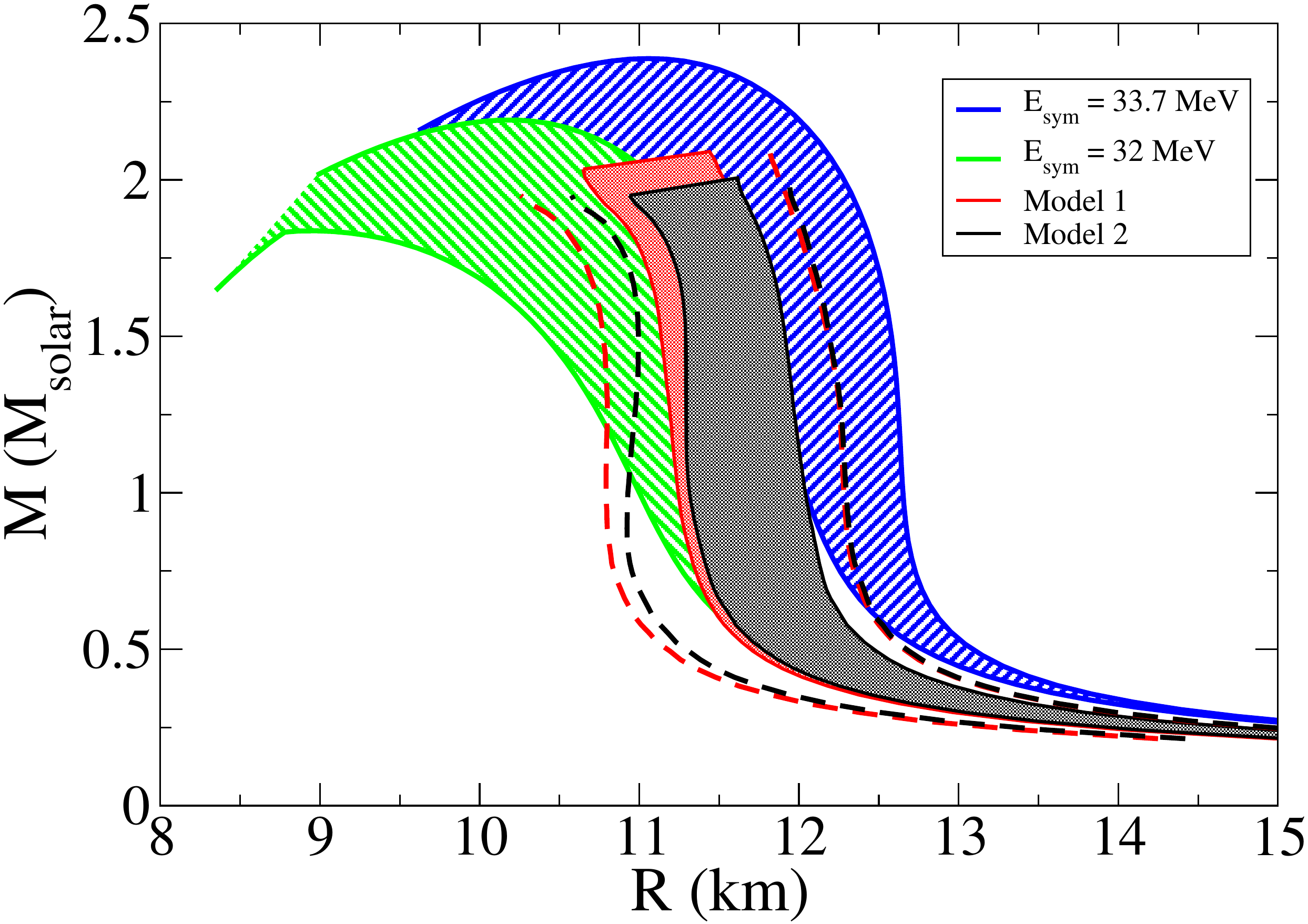}
\end{center}
\caption{The comparison of $M-R$ relation of neutron stars obtained 
from QMC calculations and observations.
The blue and green bands are the same of Fig.~\ref{fig:nstar}, and
correspond to EOS giving the value of $E_{sym}$ indicated in the legend.
The black and red bands are obtained from neutron star observations of
Ref.~\cite{Steiner:2012} at the $1$-$\sigma$ confidence level (dashed lines
at $2$-$\sigma$), and they correspond to different models
of the high-density part.
}
\label{fig:mvsr}
\end{figure}

In order to better appreciate the agreement between theoretical
calculation with the neutron star structure obtained from observations,
we show a comparison in Fig.~\ref{fig:mvsr}. In the figure the two
green and blue bands correspond to the $M-R$ relation obtained from the
EOS of Fig.~\ref{fig:eos}, and the black and red bands represent the
astrophysical observation of Ref.~\cite{Steiner:2012} using different
models for the high-density EOS.

\section{Conclusions}

We have presented a theoretical calculation of the neutron matter EOS
using QMC methods. This technique permits to study the ground-state of
strongly interacting Fermi systems in a full non-perturbative way.
Calculations have been performed using a modern nucleon-nucleon 
interaction that fit the phase shifts with high accuracy.
We have studied the effect of using different microscopical models 
of three-neutron forces, by quantifying their role in the high-density
EOS up to 3$\rho_0$. By performing simulations using Hamiltonians that
give different values of the symmetry energy we conclude that, at present,
the uncertainty to the EOS is mainly due to the poor constrain of $E_{sym}$ 
rather than the model of the three-neutron force.
From our calculation we have extracted the relation between $L$ and $E_{sym}$
suggesting that they are quite strongly linearly related.

We also provide new constraints from astrophysical observations. 
By combining the recent analysis of Steiner {\emph et al.} with an
empirical EOS which form is suggested by QMC simulations, we provided
a new constrain to the value of the symmetry energy and its slope at
saturation~\cite{Steiner:2012}. The result is compatible with several experimental 
measurements~\cite{Tsang:2012}.
We find good agreement for the $M-R$ relation of neutron stars given 
by QMC prediction and from observations.

\ack{ 
The author would like to thank J. Carlson for critical comments on the
manuscript.  This work is supported by DOE Grants No. DE-FC02-07ER41457
(UNEDF SciDAC) and No. DE-AC52-06NA25396, and by the LANL LDRD program.
Computer time was made available by Los Alamos Open Supercomputing, and
by the National Energy Research Scientific Computing Center (NERSC).  }

\section*{References}

\bibliographystyle{iopart-num}

\providecommand{\newblock}{}

\end{document}